\documentclass[]{aa} 
\usepackage{txfonts}
\usepackage{graphicx}

\begin{document}

\title{Binaries discovered by the SPY project}
\subtitle{
V. GD\,687 -- a massive double degenerate binary progenitor that will merge within a Hubble time
\thanks{Based on observations at the Paranal Observatory of the European 
Southern Observatory for programme No. 165.H-0588(A). Based on observations at 
the La Silla Observatory of the European Southern Observatory for programmes
 No. 072.D-0510(B), 079.D-0288(A), 080.D-0685(A) and 084.D-0348(A).}
}

\author{S. Geier \inst{1}
   \and U. Heber \inst{1}
   \and T. Kupfer \inst{1}
   \and R. Napiwotzki \inst{2}}

\offprints{S.\,Geier,\\ \email{geier@sternwarte.uni-erlangen.de}}

\institute{Dr.\,Karl Remeis-Observatory \& ECAP, University Erlangen-Nuremberg, Sternwartstr. 7, 96049 Bamberg, Germany 
   \and Centre of Astrophysics Research, University of Hertfordshire, College
    Lane, Hatfield AL10 9AB, UK}

\date{Received \ Accepted}

\abstract{
{\it Aims.} The ESO SN Ia Progenitor Survey (SPY) aims at finding merging 
double degenerate binaries as candidates for supernova type Ia (SN Ia) explosions. A 
white dwarf merger has also been suggested to explain the formation of rare types of stars like R CrB, extreme helium or He sdO stars. Here we present the hot subdwarf B binary GD\,687, which will merge in less than a Hubble time.\\  
{\it Methods.} The orbital parameters of the close binary have been determined 
from time resolved spectroscopy. Since GD\,687 is a  single-lined binary, 
the spectra contain only information about the subdwarf primary and its orbit. 
From high resolution spectra the projected rotational velocity was derived.
Assuming orbital synchronisation, the inclination of the system and the mass of the unseen companion were constrained.\\ 
{\it Results.} The derived inclination is $i=39.3^{+6.2}_{-5.6}\,^{\circ}$. The mass $M_{\rm 2}=0.71_{-0.21}^{+0.22}\,M_{\rm \odot}$ indicates that the companion must be a white dwarf, most likely of C/O composition. This is only the fourth case that an sdB companion has been proven to be a white dwarf unambiguously. Its mass is somewhat larger than the average white dwarf mass, but may be as high as $0.93\,M_{\rm \odot}$ in which case the total mass of the system comes close to the Chandrasekhar limit.  \\
{\it Conclusions.} GD\,687 will evolve into a double degenerate system and merge
to form a rare supermassive white dwarf with a mass in excess of solar.
A death in a sub-Chandrasekhar supernova is also conceivable.

}

\maketitle

\section{Introduction \label{sec:intro}}

Double degenerate (DD) binaries consisting of two white dwarf (WD) stars experience a 
shrinkage of their orbits caused by gravitational wave radiation. 
Sufficiently close binaries with orbital periods of less than half a day 
will eventually merge within less than a Hubble time. The outcome of such a 
merger can be a single compact object like a white dwarf or a neutron star. 
But the merger event may also lead to the ignition of nuclear burning 
and trigger the formation of rare objects such as extreme helium stars 
(Saio \& Jeffery \cite{saio}), R CrB stars (Webbink \cite{webbink}) or He sdO stars (Heber \cite{heber3}; Napiwotzki \cite{napiwotzki10}). 
%Justham et al. \cite{justham}
The merger of two sufficiently massive C/O white dwarfs may lead to a 
Supernova of type Ia (SN~Ia, Webbink, \cite{webbink}). 

SN~Ia play a key role in the study of cosmic 
evolution. They are utilised as standard candles for determining the 
cosmological parameters (e.g. Riess et al. \cite{riess}; 
Leibundgut \cite{leibundgut}; Perlmutter et al. \cite{perlmutter}). 
There is general consensus that the thermonuclear explosion of a 
white dwarf of Chandrasekhar mass causes a SN~Ia. 
The DD merger (Iben \& Tutukov \cite{iben}) is one of two 
 main scenarios to feed a white dwarf to the Chandrasekhar mass.
Progenitor candidates for the DD scenario have to merge in less than a Hubble time and the total binary mass must exceed the Chandrasekhar limit. 

In general, DD candidates are assumed to consist of two white dwarfs. 
The discovery of the sdB+WD binary KPD\,1930$+$2752, however, has highlighted 
the importance of such systems as SN~Ia progenitor candidates 
(Bill\`{e}res et al. \cite{billeres}; Maxted et al. \cite{maxted1}; Geier et al. \cite{geier1}). 
Hot subdwarf stars (sdBs) are core helium-burning stars situated at the hot 
end of the horizontal branch with masses around $0.5\,M_{\rm \odot}$ 
(Heber \cite{heber1}). Although the formation of these objects is still 
under debate, it is general consensus that they must loose most of their 
envelope at the tip of the RGB and evolve directly from the EHB to the 
WD cooling tracks without ascending the AGB (see Heber \cite{heber} for a review). 

Since a high fraction of sdBs resides in close binaries 
with unseen companions (Maxted et. al \cite{maxted2}; Napiwotzki et al. \cite{napiwotzki6}), such systems can qualify as SN~Ia progenitors, if the companion is a white dwarf and the 
lifetime of the sdB is shorter than the merging time of the binary. This is the 
case for the sdB+WD binary KPD\,1930+2752, which is the best known DD progenitor
 candidate for SN~Ia (Maxted et al. \cite{maxted1}; Geier et al. \cite{geier1}). 

Systematic radial velocity (RV) searches for DDs have been undertaken 
(e.g. Napiwotzki \cite{napiwotzki2} and references therein). 
The largest of these projects was the ESO SN~Ia Progenitor Survey (SPY). 
More than $1000$ WDs were checked for RV-variations 
(Napiwotzki et al. \cite{napiwotzki9,napiwotzki2}). 
SPY detected $\sim 100$ new DDs (only $18$ were known before). 
One of them may fulfil the criteria for SN~Ia progenitor candidates 
(Napiwotzki et al. \cite{napiwotzki3}).

This sample includes about two dozen radial velocity variable sdB stars with
invisible companions. As the sdB stars are intrinsically brighter than the white
dwarfs, the nature of the companion is not so easily revealed.  

Most sdBs in close binary systems are single-lined and no features of the 
companions are visible in their spectra. In this case the nature of the unseen 
companion is hard to constrain since only lower limits can be derived for the 
companion masses. Given the fact that sdBs are subluminous stars, main 
sequence companions earlier than K-type can easily be excluded. 
But if the derived lower limit for the companion mass is lower than 
about half a solar mass it is in general not possible to distinguish between a  main sequence companion and a compact object like a white dwarf. 
Despite the fact that the catalogue of Ritter \& Kolb (\cite{ritter}) lists more
 than 80 sdB binaries, the nature of their companions can only be 
constrained in special cases (e.g. For et al. \cite{for}; Geier et al. 
\cite{geier2}), in particular from photometric variations due to eclipses, reflection effects or ellipsoidal deformations. 

Results for eight DD systems discovered in the SPY survey have been presented 
in papers I--IV (Napiwotzki et al. \cite{napiwotzki5}; Napiwotzki et al. 
\cite{napiwotzki7}; Karl et al. \cite{karl2}; Nelemans et al. \cite{nelemans}). 
Here we study the sdB binary GD\,687, a 
DD progenitor system which will merge in less than a Hubble time.

\begin{figure}[t!]
	\centering
	\resizebox{\hsize}{!}{\includegraphics{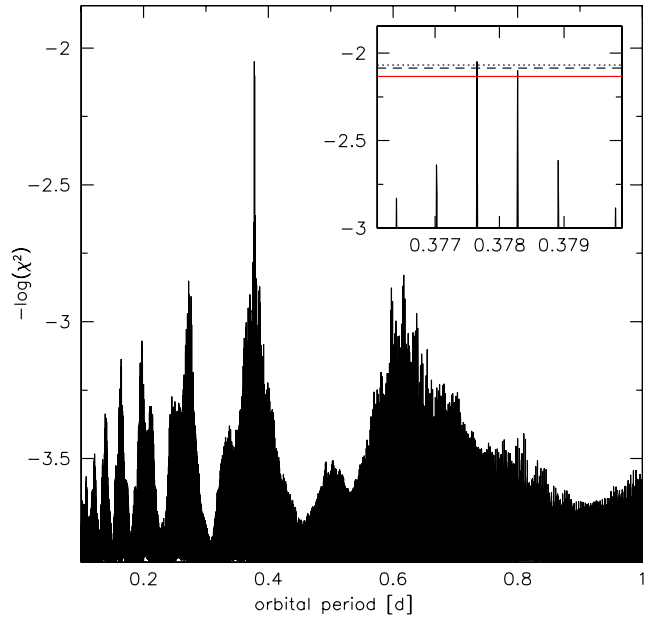}}
	\caption{In this power spectrum $-\log\chi^{2}$ of the best sine fit is plotted against the orbital periods. The region around the best solution is shown in the inlet. Confidence limits ($1\sigma$,$3\sigma$,$6\sigma$) are marked with horizontal lines (dotted, dashed, solid).}
	\label{power}
\end{figure}

\section{Binary parameters \label{sec:par}}

\subsection{Observations and radial velocity curve}

GD\,687 ($\alpha_{2000}=01^{\rm h}10^{\rm m}18{\stackrel{\rm s}{\displaystyle 
.}}5$, $\delta_{2000}=-34^{\rm \circ}00'26''$) was included in the SPY survey, 
because it was misclassified as DA3 white dwarf (McCook \& Sion \cite{mccook}).
 GD\,687 was observed twice in the course of the SPY project with the high
  resolution echelle spectrograph UVES at the ESO\,VLT. 
The spectra have a resolution of $R\approx37\,000$ and cover $3200-6650\,{\rm \AA}$ with two small gaps at $4580\,{\rm \AA}$ and $5640\,{\rm \AA}$.

Follow-up medium resolution spectra were taken with the EMMI ($R\approx3400,\lambda=3880-4380\,{\rm \AA}$) and the EFOSC2 ($R\approx2200,\lambda=4450-5110\,{\rm \AA}$) spectrographs mounted at the ESO\,NTT. Reduction was done with the ESO--MIDAS package. The radial velocities (RV) were measured by fitting a set of mathematical functions (Gaussians, Lorentzians and polynomials) to the hydrogen Balmer lines using the FITSB2 routine (Napiwotzki et al. \cite{napiwotzki4}). The measured RVs are given in Table~\ref{RVs}. 

Sine curves were fitted to the RV data points in fine steps over a range of 
test periods. For each period the $\chi^{2}$ of the best fitting sine curve 
was determined. The result is similar to a power spectrum with the lowest 
$\chi^{2}$ indicating the most likely period (see Fig.~\ref{power}). 
The plot shows two very closely spaced peaks. We performed a Monte Carlo
simulation for the most likely periods. For each iteration a
randomised set of RVs was drawn from Gaussian distributions with
central value and width corresponding to the RV measurements and the
analysis repeated.

The probability that the solution with the lowest $\chi^{2}$ and 
$P=0.37765\,{\rm d}$ is the correct one is estimated to be $74\%$. 
The second best alias period ($P=0.37828\,{\rm d}$) has $\Delta\chi^{2}=9$ and 
a probability of $26\%$ to be the correct one (see Fig.~\ref{power} inlet). None of the $10\,000$ Monte Carlo iterations indicated a period different from the two discussed above. The probability that any other solution is the correct one is therefore less than $0.01\%$. Our analysis is based on the most likely solution. 

\begin{figure}[t!]
	\centering
	\resizebox{\hsize}{!}{\includegraphics{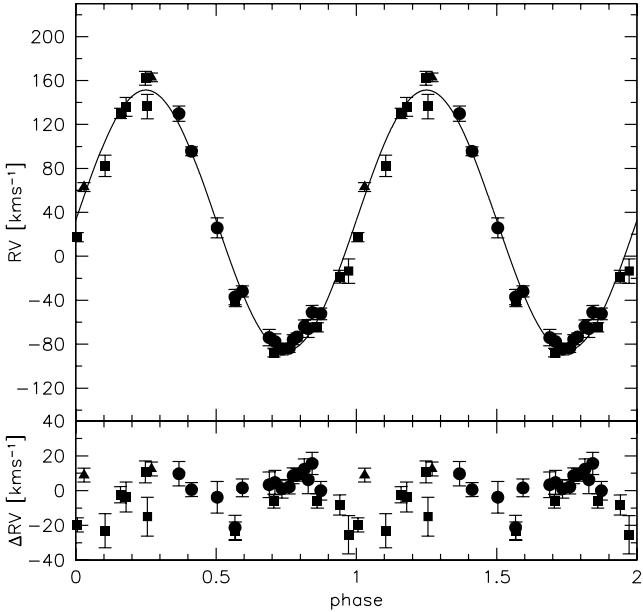}}
	\caption{Radial velocities of the primary subdwarf plotted against 
	orbital phase. The residuals are plotted below. Filled triangles mark 
	RVs measured from UVES spectra, filled diamonds mark RVs obtained from 
	EMMI spectra and filled circles RVs measured from EFOSC2 spectra.}
	\label{rv}
\end{figure}

In order to derive most conservative errors for the RV semi-amplitude $K$ and 
the system velocity $\gamma$ we fixed the most likely period, created new RV 
datasets with a bootstrapping algorithm and calculated the orbital solution in 
each case. The standard deviation of these results was adopted as error 
estimate and is about three times higher than the $1\sigma$-error. 
The phase folded RV curve is shown in Fig.~\ref{rv}. Parameters are given in 
Table~\ref{tab:par}\footnote{Adopting the second best solution only leads to small  changes in the derived parameters ($K=124.2\pm3.0\,{\rm km\,s^{-1}}$,
$\gamma=37.9\pm3.2\,{\rm km\,s^{-1}}$) and has no significant impact on the 
qualitative discussion of our results.}.

\begin{table}
\caption{Radial velocities of GD\,687}
\label{RVs}
\begin{center}
\begin{tabular}{lrl}
\hline
\noalign{\smallskip}
mid$-$HJD & RV [${\rm km\,s^{-1}}$] & Instrument\\
\noalign{\smallskip}
\hline
\noalign{\smallskip}
2451737.83523 & 162.9 $\pm$ 4.0 & UVES \\ 
2451740.76555 & 63.1  $\pm$ 4.0 &  \\
\noalign{\smallskip}
\hline
\noalign{\smallskip}
2453337.67383 & -40.5 $\pm$ 5.3 & EMMI \\
2453338.59413 &	17.4 $\pm$ 4.1 &      \\
2453339.61394 &	-87.6 $\pm$ 4.0	&      \\
2453339.67229 & -64.7 $\pm$ 4.1 &      \\
2453340.54050 & 130.1 $\pm$  4.7 &      \\
2454252.87242 & -13.5 $\pm$  11.0 &      \\
2454252.92242 &  82.4  $\pm$  9.8 &      \\
2454254.86756 &  136.3 $\pm$  11.1 &      \\
2454476.54549 &  162.1 $\pm$  6.3 &      \\
2454477.56241 &  -18.6 $\pm$  5.8 &      \\
2454477.65241 &  136.1 $\pm$  8.6 &      \\
\noalign{\smallskip}
\hline
\noalign{\smallskip}
2455144.65365 &   129.9 $\pm$  7.0  & EFOSC2 \\
2455144.67058 &    95.6 $\pm$  4.0  & \\
2455144.70520 &    25.9 $\pm$  9.1  & \\
2455144.72910 &   -37.1 $\pm$  7.0  & \\
2455144.73909 &   -31.9 $\pm$  5.1  & \\     
2455144.77512 &   -74.0 $\pm$  7.4  & \\     
2455145.53816 &   -77.6 $\pm$  7.1  & \\     
2455145.54772 &   -84.4 $\pm$  5.4  & \\    
2455145.55739 &   -83.7 $\pm$  4.0  & \\    
2455145.56266 &   -76.0 $\pm$  4.5  & \\     
2455145.56792 &   -73.4 $\pm$  3.6  & \\     
2455145.57774 &   -63.9 $\pm$  5.9  & \\     
2455145.58299 &   -65.8 $\pm$  8.0  & \\    
2455145.58826 &   -51.1 $\pm$  6.4  & \\     
2455145.59991 &   -52.4 $\pm$  5.3  & \\     
\noalign{\smallskip}
\hline

\end{tabular}
\end{center}
\end{table}

\subsection{Gravity and projected rotational velocity \label{sec:ana}}
 
The atmospheric parameters of GD\,687 were measured by fitting metal-line 
blanketed LTE models to the UVES spectra and are given in Lisker et al. (\cite{lisker}). 

In order to derive $v_{\rm rot}\,\sin{i}$ and the elemental abundances, we 
coadded the observed high resolution spectra after shifting them to rest 
wavelength. The spectrum was then compared with rotationally broadened, 
synthetic line profiles calculated with the LINFOR program (developed by 
Holweger, Steffen and Steenbock at Kiel university, modified by Lemke 
\cite{lemke}). 

Due to the wide slit used ($2.1''$) for the SPY survey (see Napiwotzki et al. 
\cite{napiwotzki9} for details) the resolution of the UVES spectra is seeing 
dependent in most cases. In order to measure  $v_{\rm rot}\,\sin{i}$ 
accurately, the instrumental profile has to be taken into account. We used the 
ESO archive to obtain the seeing conditions during the exposure times of the 
UVES spectra (DIMM seeing monitor, Sarazin \& Roddier \cite{sarazin}), took 
the average ($1.13\,{\rm arcsec}$) and folded our models with the instrumental 
profile.
 
The projected rotational velocity was measured simultaneously with 
the elemental abundances to $v_{\rm rot}\sin{i}=21.2\pm2.0\,{\rm km\,s^{-1}}$ 
using the six strongest metal lines. In addition, an abundance and rotational broadening fit was done using two 
prominent helium lines (He\,{\sc i} 4472, 5876) and LTE model spectra 
(Heber et al. \cite{heber2}). The resulting 
$v_{\rm rot}\sin{i}=19.8\pm2.2\,{\rm km\,s^{-1}}$ turned out to be perfectly 
consistent with the value measured from the metal lines. Since metal lines 
are more sensitive to rotational broadening we adopt the value derived from 
these lines for our analysis. The parameters are given in Table~\ref{tab:par}.

Our results depend very much on the accuracy of the $v_{\rm rot}\sin{i}$ and 
$\log{g}$ measurements. A thorough discussion of the error in surface gravity 
is given by Lisker et al. (\cite{lisker}). To quantify the 
$v_{\rm rot}\sin{i}$ error, we carried out numerical simulations. For this, 
synthetic spectra with fixed rotational broadening were computed and convolved 
with the instrumental profile. Random noise was added to mimic the observed 
spectra. The rotational broadening was measured in the way described above 
using a grid of synthetic spectra for various rotational and instrumental 
broadenings as well as S/N levels. Variations in the instrumental profile  and 
the noise level were the dominant error sources. 
%The average errors are of the order of typically $1.0-3.0\,{\rm kms^{-1}}$. 
Behr (\cite{behr}) used a similar method to measure the low $v_{\rm rot}\sin{i}$ of Blue Horizontal 
Branch stars from high resolution spectra. The errors given in that work are of the same 
order as the one given here for GD\,687 ($2.0\,{\rm km\,s^{-1}}$).

\begin{figure}[t!]
	\centering
	\resizebox{\hsize}{!}{\includegraphics{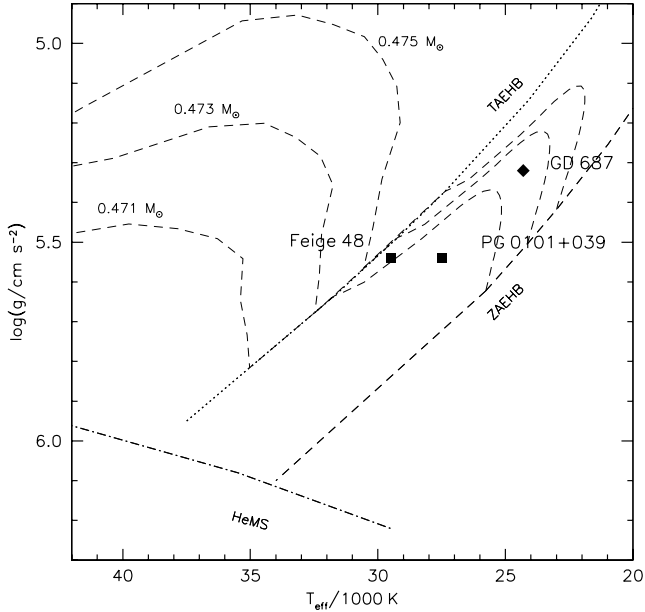}}
	\caption{$T_{\rm eff}$-$\log{g}$-diagram. The location of the EHB and the evolutionary tracks (dashed curves) are taken from Dorman et al. (\cite{dorman}).}
	\label{tefflogg}
\end{figure}

\subsection{Analysis \label{sec:analysis}}

The analysis strategy used is described only briefly. For details we refer
the reader to Geier et al. (\cite{geier1}, \cite{geier2}). Since the spectrum 
of GD\,687 is single-lined, it 
contains no information about the orbital motion of the companion, 
and thus only the mass function 
$f_{\rm m} = {M_{\rm comp}^3 \sin^3i}/(M_{\rm comp} + M_{\rm sdB})^2 = 
P K^3/2 \pi G$ can be calculated. Although the RV semi-amplitude $K$ and the 
period $P$ are determined
 by the RV curve, $M_{\rm sdB}$, $M_{\rm comp}$ and $\sin^3i$ remain
 free parameters.

Nevertheless, the masses can be constrained by assuming tidal
synchronisation (see also Napiwotzki et al. \cite{napiwotzki5}). Combining the 
orbital parameters with an estimate of the sdB mass and with the 
determination of its $v_{\rm rot}\sin{i}$ and surface 
gravity, allows the mass of the invisible companion to be constrained. 
The mass of the sdB primary is taken from the population synthesis 
models (Han et al. \cite{han1}, \cite{han2}) 
that predict a mass range of $M_{\rm sdB}$\,=\,0.37 -- 0.48\,M$_{\rm \odot}$ 
for sdBs in binaries, which experienced a common envelope ejection. 
The mass distribution shows a sharp peak at a mass of 
about $0.47\,{\rm M_{\odot}}$ (see Fig.~22 of Han et al. \cite{han2}) ranging
from 0.43 to 0.47\,M$_{\rm \odot}$. This theoretical mass 
distribution is consistent with analyses of close 
binary systems (e.g. Geier et al. \cite{geier1}) as well as asteroseismic 
analyses of pulsating sdBs (see Charpinet et al. \cite{charpinet} and references therein). 

If the rotational period of the sdB primary is synchronised the rotational velocity 
$v_{\rm rot}= 2 \pi R_{\rm sdB}/P$ can be calculated. 
The radius of the primary is given by the mass radius relation 
$R = \sqrt{M_{\rm sdB}G/g}$. The measurement of the 
projected rotational velocity $v_{\rm rot}\,\sin\,i$ therefore allows us to 
constrain the inclination angle $i$.
For the most likely sdB mass $M_{\rm sdB}=0.47\,{\rm M_{\odot}}$ the mass 
function can be solved, and both the inclination angle and the companion 
mass can be derived. The errors are calculated by chosing the most extreme values for the input 
parameters within their respective error limits. In order to account for the theoretical uncertainity 
in sdB mass, we adopted the predicted mass range for the sdB ($0.43-0.47\,{\rm M_{\odot}}$) and calculated the lower limit for 
the companion mass under the assumption that $M_{\rm sdB}=0.43\,{\rm M_{\odot}}$. 
The error budget is dominated by the uncertainties in the 
$v_{\rm rot}\sin{i}$ and $\log{g}$ measurements. 
%The errors in $P$ and $K$ 
%turned out be negligible.

\begin{table}[h!]
\caption{Orbital and atmospheric parameters of GD\,687. \dag Taken from Lisker 
et al. (\cite{lisker}). \ddag The merging time is calculated using the formula given in Ergma et al. (\cite{ergma}).} 
\label{tab:par}
\begin{center}
\begin{tabular}{ll}
	\hline
        \noalign{\smallskip}
        Orbital parameters & \\
        \noalign{\smallskip}
        \hline
        \noalign{\smallskip}
        $T_{\rm 0}$ [HJD]    & $2455144.515\pm0.002$ \\
        $P$                  & $0.37765\pm0.00002\,{\rm d}$ \\
        $\gamma$             & $32.3\pm3.0\,{\rm km\,s^{-1}}$ \\
        $K$                  & $118.3\pm3.4\,{\rm km\,s^{-1}}$ \\
        $f(M)$               & $0.065\pm0.006\,{\rm M_{\odot}}$ \\
        \hline
        \noalign{\smallskip}
        Atmospheric parameters & \\
        \noalign{\smallskip}
        \hline
        \noalign{\smallskip}
        $T_{\rm eff}$        & $24350\pm360\,{\rm K}$\dag \\
        $log\,g$             & $5.32\pm0.07\,{\rm K}$\dag \\
        $v_{\rm rot}\sin{i}$ & $21.2\pm2.0\,{\rm km\,s^{-1}}$\\
        \hline
        \noalign{\smallskip}
        Derived binary parameters & \\
        \noalign{\smallskip}
        \hline
        \noalign{\smallskip}
        $M_{\rm 1}$ (adopted) & $0.47\,{\rm M_{\odot}}$ \\
        $R_{\rm 1}$           & $0.25\pm0.02\,{\rm R_{\odot}}$ \\
        $i$                   & $39.3^{+6.2}_{-5.6}\,^{\circ}$ \\
        $M_{\rm 2}$           & $0.71_{-0.21}^{+0.22}\,{\rm M_{\odot}}$ \\
        $M_{\rm 1}+M_{\rm 2}$ & $1.18_{-0.21}^{+0.22}\,{\rm M_{\odot}}$ \\
        $t_{\rm merger}$      & $11.1\times10^{9}\,{\rm yr}$\ddag \\
	\hline
\end{tabular}
\end{center}
\end{table}

\section{Nature of the unseen companion}

The mass function provides a lower limit to the mass of the invisible companion
of $0.35\,{\rm M_{\odot}}$. In the case of a white dwarf primary it is
impossible to hide the contribution of a main sequence star even of the lowest
mass in optical/NIR spectra since these are intrinsically faint. This is not the case for sdB stars.
A main sequence companion with a mass lower than $0.45\,{\rm M_{\odot}}$ can not 
be excluded because its luminosity would be too low to be 
detectable in the spectra (Lisker et al. \cite{lisker}).
This is the reason why the companions' nature still remains unknown for most of
the $\approx$80 sdB systems in the catalogue of Ritter \& Kolb (\cite{ritter}).
Additional information is needed. 

No spectral features of a cool main sequence star are present in the optical
spectra of GD\,687. Furthermore, Farihi et al. 
(\cite{farihi}) included GD\,687 in a near-infrared imaging survey to search 
for low-luminosity companions to white dwarfs and found no evidence for an 
infrared excess which could be caused by a cool main sequence companion. 

As the lower limit for the mass of GD\,687's companion derived from the mass
function is lower than 
$0.45\,{\rm M_{\odot}}$, additional information is needed to clarify its 
nature. We made use of the gravity and projected rotational velocity to
constrain the mass to $0.71_{-0.21}^{+0.22}\,{\rm M_{\odot}}$.  
The companion therefore can not be a main sequence star but has to be a white
dwarf. Taking into 
account the possible mass range it is very likely to be of C/O composition.
Its mass exceeds that of an average white dwarf.

GD\,687 is only the fourth sdB star, for which the white dwarf nature of the companion could be shown unamibiguously. The others have been discovered by analysing ellipsoidal light variations (KPD\,1930$+$2752, 
PG\,0101$+$039) or eclipses (KPD\,0422$+$5421, Orosz \& Wade \cite{orosz}). KPD\,1930$+$2752 and PG\,0101$+$039 have been confirmed with the method used here (Geier et al. \cite{geier1,geier2}).

\section{Discussion}

The derived companion mass was calculated under the assumption of orbital
 synchronisation. Since theoretical synchronisation timescales for hot stars 
 with radiative envelopes are not consistent (Zahn \cite{zahn}; 
 Tassoul \& Tassoul \cite{tassoul}), empirical evidence for orbital 
 sychronisation in sdB binaries is needed. Geier et al. (\cite{geier2}) 
 found such evidence by detecting a variation in the lightcurve of the sdB+WD 
 binary PG\,0101$+$039, which could be identified as ellipsoidal deformation of 
 the sdB. Since the orbital period of PG\,0101$+$039 is $0.57\,{\rm d}$, sdB 
 binaries with shorter periods like GD\,687 are very likely synchronised as 
 well. Recently van Grootel et al. (\cite{vangrootel}) performed an 
 asteroseismic analysis of the pulsating sdB binary Feige 48 and for the first 
 time proved orbital sychronisation in this way. The orbital period of Feige 48 
 ($0.36\,{\rm d}$) is very similar to the one of GD\,687. Furthermore, the 
 atmospheric parameters of GD\,687 ($T_{\rm eff}=24\,300\,{\rm K}$,
$\log{g}=5.32$) indicate that it has already evolved away from the ZAEHB
(Zero Age Extreme Horizontal Branch) and should therefore, according to
evolutionary calculations (e.g. Dorman et al. \cite{dorman}), have a similar age as PG\,0101$+$039 and Feige\,48 (see Fig.~\ref{tefflogg}). We thus conclude that the assumption of orbital sychronisation is fully justified in the case of GD\,687.

In about $100\,{\rm Myr}$ the helium-burning in the core of the sdB will come to an end. After a short period of helium-shell-burning this star will eventually become a white dwarf consisting of C and O. GD\,687 is one of only a few known DD progenitor systems, where both components are C/O white dwarfs and which will merge in less than a Hubble time.

\begin{figure}[t!]
	\resizebox{\hsize}{!}{\includegraphics{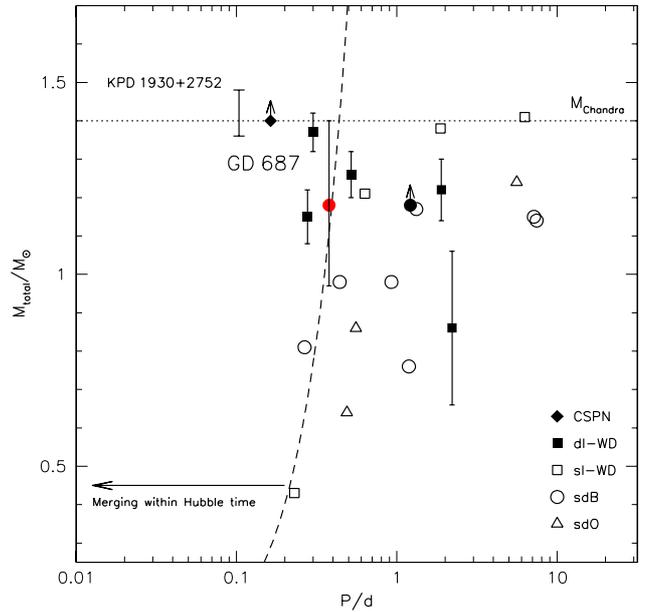}}
	\caption{Total mass plotted against logarithmic period of double 
degenerate systems from the SPY survey. GD\,687 is marked with the
filled circle. For KPD\,1930$+$2752 a mass range is given 
(Geier et al. \cite{geier1}). The filled rectangles mark double-lined 
WDs, for which the absolute masses can be derived. The filled circle 
with arrow marks the lower mass limit derived for HE\,1047$-$0436 
(Napiwotzki et al. \cite{napiwotzki5}), the filled diamond the lower 
mass limit derived for PN\,G\,135.9$+$55.9 (Napiwotzki et al. 
\cite{napiwotzki8}). The open symbols mark single-lined WDs, sdBs, and 
sdOs. The companion masses of the single-lined systems are derived 
for the expected average inclination angle ($i=52^{\circ}$) 
(Napiwotzki et al. \cite{napiwotzki7}; Karl et al. \cite{karl2}; 
Karl \cite{karl}; Nelemans et al. \cite{nelemans}; Napiwotzki et 
al. \cite{napiwotzki3}).
}
\label{progen}
\end{figure}

Compared to the sdB+WD binary KPD\,1930$+$2752 ($P\approx0.1\,{\rm d}$, Maxted 
et al. \cite{maxted1}; Geier et al. \cite{geier1}), which is the best known 
candidate for DD SN~Ia progenitor, the orbital period of GD\,687 is rather 
long. This leads to a merging time of $11.1\,{\rm Gyr}$, which is just a 
little shorter than the Hubble time, compared to only $200\,{\rm Myr}$ for 
KPD\,1930$+$2752. With a total mass of $1.18_{-0.21}^{+0.22}\,M_{\rm \odot}$ 
for the most likely subdwarf mass it may come close to the Chandrasekhar limit of 
$1.4\,M_{\rm \odot}$ and is therefore placed at the edge of the progenitor parameter space (see Fig. \ref{progen}). In contrast to KPD\,1930$+$2752, where the primary mass could be constrained by an additional analysis of the subdwarfs ellipsoidal deformation visible as variation in its lightcurve, no such 
constraint can be put on the primary mass of GD\,687 yet. 

Instead of exploding as SN~Ia, the merger of the two white dwarfs will most likely lead to the formation of a supermassive white dwarf with O/Ne/Mg-core.

Up to now four binaries with total masses between about $1.2$ and 
$1.4\,{\rm M_{\odot}}$ have been discovered. In two of them the visible 
component is an sdB. Since the 
sdB close binary fraction is much higher than the one of white dwarfs, 
it may be easier 
to find double-degenerate binary progenitors in the hot subdwarf population. 

Two other massive DD systems, a central star of a 
planetary nebula (Tovmassian et al. \cite{tovmassian}; Napiwotzki et al. \cite{napiwotzki8})
and a white dwarf from the SDSS survey (Badenes et al. \cite{badenes}; Marsh et al. \cite{marsh}; Kulkarni \& van Kerkwijk \cite{kulkarni}), were discovered serendipitously.

Even though GD\,687 does not qualify as SN~Ia progenitor candidate, 
the discovery of a population of 
double degenerate binaries (and progenitor systems) with total masses close to the Chandrasekhar limit 
in the course of the SPY survey (see Fig. \ref{progen}) provides evidence for 
a similar population exceeding this limit. The same binary evolution channel 
that produces the sub-Chandrasekhar systems will also produce 
super-Chandrasekhar systems with slight changes in the initial conditions only. 

\begin{acknowledgements}

S.~G. is supported by the Deutsche Forschungs\-gemeinschaft under grant 
He1354/40-3.

\end{acknowledgements}

\end{document}